\documentclass[12pt]{article}

 \def\lddots{\mathinner{\mkern1mu\raise1pt\hbox{.}\mkern2mu

    \raise4pt\hbox{.}\mkern2mu\raise7pt\vbox{\kern7pt\hbox{.}}\mkern1mu}}

\makeatletter

\def\numberbysection{\@addtoreset{equation}{section}

    \def\theequation{\thesection.\arabic{equation}}}

\makeatother

\numberbysection

\newcommand{\be}{\begin{eqnarray}}

\newcommand{\ee}{\end{eqnarray}}

\newcommand{\non}{\nonumber}

\newcommand{\tr}{\mathop{\rm tr}\nolimits}

\begin{document}

\begin{titlepage}

\strut\hfill

\vspace{.5in}

\begin{center}

\LARGE  The $XXX$ spin $s$ quantum chain and the alternating
$s^{1}$, $s^{2}$ chain with boundaries\\[1.0in]

\large  Anastasia Doikou \footnote{e-mail: ad22@york.ac.uk} \\

\normalsize{ Department of Mathematics, University of York,
Heslington\\ York YO10 5DD, United Kingdom}\\

\end{center}

\vspace{.5in}

\begin{abstract}

The integrable $XXX$ spin $s$ quantum chain and the alternating
$s^{1}$, $s^{2}$ ($s^{1}-s^{2}={1\over 2}$) chain with boundaries
are considered. The scattering of their excitations with the
boundaries via the Bethe ansatz method is studied, and the exact
boundary $S$ matrices are computed in the limit $s, s^{1, 2} \to
\infty$. Moreover, the connection of these models with the $SU(2)$
Principal Chiral, $WZW$ and the $RSOS$ models is discussed.

\end{abstract}

\end{titlepage}

\section{Introduction and Review}

 Quantum spin chains are 1D quantum mechanical models
of $N$ microscopic degrees of freedom, namely spins, with the
Heisenberg ($XXX$) model solved by Bethe \cite{bethe}, \cite{FT2},
\cite{FT1} being the prototype. Two types of quantum spin chains
exist, known as ``closed'' (periodic boundary conditions) and
``open''. Closed spin chains have been very well studied (see e.g.
\cite{FT2}--\cite{R}), whereas open chains are less widely
studied, although of great interest. A very interesting aspect of
these models is that in the continuum limit, they correspond to
1+1--dimensional integrable quantum field theories \cite{affl}.
 For example the $S$ matrix that describes the anisotropic Heisenberg
 ($XXZ$) model in a
certain regime also describes a massive integrable quantum field
theory, namely the Thirring model or the sine--Gordon model. Also,
the $S$ matrix that describes the bulk scattering for the spin $s$
$XXX$ chain coincides with the one of the
Wess--Zumino--Witten ($WZW$) model \cite{resh2}.
 Finally, open spin chains and the
corresponding 2D field theories with boundaries display a rich
variety of boundary phenomena which --- for integrable models ---
can be investigated exactly.

During the last years, there has been an increased research
interest on integrable models with boundaries, especially after
the prototype work of Cherednik and Sklyanin \cite{cherednik},
\cite{sklyanin}. In particular, Cherednik introduced the
reflection equation in order to obtain boundary scattering
matrices for theories in half line. In the presence of boundaries
in addition to the bulk scattering, the scattering of the
particles with the boundaries ---described by the boundary $S$
matrix--- should also be considered. The boundary $S$ matrix
satisfies a collection of algebraic constraints, namely the
boundary Yang--Baxter (reflection) equation \cite{cherednik}.
Sklyanin considered this equation in the spin chain framework and
he generalized the Quantum Inverse Scattering Method for
integrable models with boundaries. Moreover, Fring and
K$\ddot{o}$berle \cite{fk} obtained solutions of the reflection
equation for the affine Toda field theories with boundaries,
whereas Ghoshal and Zamolodchikov \cite{GZ} found solutions of the
reflection equation for the sine--Gordon model on a half line.
Finally, De Vega and Gonzalez--Ruiz made similar calculations for
the $XXX$ ($SU(2)$), $XXZ$ ($A_{1}^{(1)}$), $XYZ$ and any $SU(n)$,
$A_{n-1}^{(1)}$ open spin chain \cite{DVGR3}.

In this study the integrable $XXX$ spin $s$ quantum chain and the
alternating $s^{1}$, $s^{2}$ chain with boundaries
---obtained by fusion--- are explored. As a warm up, both models
with periodic boundaries conditions are reviewed and their
relation with 2D quantum field theories is discussed.

To describe the models it is necessary to introduce the basic
constructing element, namely the $R$ matrix, which is a solution
of the Yang--Baxter equation \cite{baxter}, \cite{korepin} \be
R_{12}(\lambda_{1} - \lambda_{2})\ R_{1 3}(\lambda_{1})\
R_{23}(\lambda_{2}) = R_{23}(\lambda_{2})\ R_{1 3}(\lambda_{1})\
R_{1 2}(\lambda_{1} - \lambda_{2})\,. \label{YBE} \ee We focus on
the special case where the $R$ matrix is related to the spin $s$
representation of $SU(2)$, obtained by fusion \cite{KR},
 \be R_{0k} =
\left(
           \begin{array}{cc}
            w_{0}+w_{3}S_{k}^{3}  &w_{-}S_{k}^{-}         \\
            w_{+}S_{k}^{+}    &w_{0}-w_{3}S_{k}^{3}  \\
                                                         \end{array}\right)\,
                                                        \label{Rmatrix1}   \ee
where $w_{0} = \lambda +{i \over 2}$, $w_{3} = w_{\pm} =i$ and
$S^{3}$, $S^{\pm}$ are the $SU(2)$ generators in the spin $s$
representation and act, in general, on a $2s+1$ dimensional space
$V=C^{2s+1}$. The generators satisfy the following commutation
relations \be \left[ S^+ \,, S^-\right] = 2S^3 \,, ~~\left[ S^3
\,, S^\pm \right] = \pm S^\pm \, \ee  and for e.g. $s=1$ they
become, \be S^3_k = \left( \matrix{1 \cr
                         & 0     \cr
                         &  &-1  \cr} \right) \,, \
~~S^+_k = {\sqrt{2}} \left( \matrix{\ &1  &\  \cr
                                          \ &\  &1  \cr
                                          \ &\  &\  \cr} \right) \,,
~~S^-_k = {\sqrt{2}} \left( \matrix{\ &\  &\  \cr
                                          1 &\  &\  \cr
                                          \ &1  &\  \cr} \right)
                                          \,.\ee
For $s={1 \over 2}$ we obtain the well known $XXX$ $R$ matrix,
 \be R_{12}(\lambda)_{j j \,, j j} &=& {(\lambda + i)} \,, \non \\
R_{12}(\lambda)_{j k \,, j k} &=& { \lambda} \,, \qquad j \ne k
\,, \non \\
R_{12}(\lambda)_{j k \,, k j} &=& {i}\,, \qquad j \ne k \,, \non \\
& & 1 \le j \,, k \le 2 \,. \label{Rmatrix} \ee

 \subsection{The $XXX$ spin $s$ quantum chain}

The $XXX$ (and $XXZ$) spin $s$ integrable chain has been
intensively studied by several authors in the bulk (see
e.g.\cite{KR}--\cite{BT}). To construct the model we derive the
transfer matrix considering the $R$ matrix related to the spin $s$
representation of $SU(2)$ (\ref{Rmatrix1}). We introduce a mass
scale in our system therefore, we derive the transfer matrix with
inhomogeneities $\Omega$,
 \be t(\lambda) = \tr_{0} T_{0}(\lambda)
\ee where \be T_{0}(\lambda) = R_{0 2N}^{2s}(\lambda-\Omega)  R_{0
2N-1}^{2s}(\lambda+\Omega)\cdots R_{0 2}^{2s}(\lambda-\Omega) R_{0
1 }^{2s}(\lambda+\Omega)\,, \label{monodromy1} \ee $R_{0i}^{2s}$
is given by (\ref{Rmatrix1}) and acts on $V_{0} \bigotimes V_{i}$.
The auxiliary space $V_{0}$ is a two dimensional space whereas the
quantum space $V_{i}$, $i=1, \ldots , 2N$, is a $2s+1$ dimensional
space. After we diagonalize the transfer matrix we find the
following eigenvalues \cite{KR} \be \Lambda_{1}^{2s}(\lambda) &=&
(\lambda -\Omega + is+{i \over 2})^{N}(\lambda +\Omega + is+{i
\over 2})^{N}\prod_{j=1}^{M}\frac{(\lambda-\lambda_{j}-{i \over 2}
)}{(\lambda-\lambda_{j}+{i \over 2})} \non\\  &+&
(\lambda-\Omega-is+{i \over 2})^{N}(\lambda+\Omega-is+{i \over
2})^{N}\prod_{j=1}^{M}\frac{(\lambda-\lambda_{j}+{3i \over 2})}
{(\lambda-\lambda_{j}+{i \over 2})}. \label{eigen1} \ee The
corresponding Bethe ansatz equations are, \be
e_{q}(\lambda_{\alpha}-\Omega)^{N}e_{q}(\lambda_{\alpha}+\Omega)^{N}=
-\prod_{\beta=1}^{M} e_2(\lambda_{\alpha}-\lambda_{\beta})\,,
\label{BAE1}\ee where $q=2s$ and \be
e_n(\lambda)=\frac{\lambda+{in\over 2}}{\lambda-{in \over 2}}.
\nonumber \ee  The BAE (\ref{BAE1}) are similar to the ones found
in \cite{PW} for the lattice analogue of the $SU(2)$ $PCM$.

In general, if we fuse the auxiliary space as well, we find that
the eigenvalues of the transfer matrix are \cite{KR},\be
\Lambda_{l}^{2s}(\lambda) &=&\sum_{k=0}^{l} a_{k}(\lambda
-\Omega)^{N}a_{k}(\lambda +\Omega)^{N} \non\\ &&
\prod_{j=1}^{M}\frac{(\lambda-\lambda_{j}-{i\over 2}
)}{(\lambda-\lambda_{j}+i(k-1)+{i\over2})}\frac{(\lambda-\lambda_{j}+
il +{i\over 2})}{(\lambda-\lambda_{j}+ik+{i\over2})},
\label{eigen0}\ee where \be a_{k}(\lambda) =
\prod_{m=k}^{l-1}(\lambda -is +{i\over 2} +i
m)\prod_{n=0}^{k-1}(\lambda +is +in +{i\over 2}). \label{a_{k}}\ee
 Let us consider the special
case where $l=2s$ and $\lambda= \pm \Omega + \lambda_{0}=\pm
\Omega-{i \over 2}(2s -1)$. Then $R^{2s}(\lambda_{o})$ becomes the
permutation operator, therefore we can derive a local Hamiltonian
for the system (up to an additive constant) \be H \propto {i\over
2 \pi} {d \over d\lambda} \log(t^{2s}(\lambda))
\vert_{\lambda=\Omega +\lambda_{0}} +{i\over 2 \pi}{d \over
d\lambda} \log(t^{2s}(\lambda)) \vert_{\lambda=-\Omega
+\lambda_{0}}. \label{hamil} \ee The corresponding eigenavalues
follow from (\ref{eigen0}) and (\ref{hamil})  \be E=-{1 \over 2
\pi }\sum_{n=1}^{2}\sum_{j=1}^{M} \ { q \over (\lambda_{j}+
(-)^{n} \Omega + {iq\over 2})(\lambda_{j}+(-)^{n} \Omega- {iq\over
2})} \label{energya} \ee also the momentum and spin are given by
\begin{equation}
P= {1 \over 2 i} \sum_{n=1}^{2}\sum_{j=1}^{M}  \log {(\lambda
_{j}+(-)^{n}\Omega+ {iq\over 2}) \over( \lambda_{j}+(-)^{n} \Omega
-{iq\over 2})}\label{EPS1}
\end{equation}
\begin{equation}
S^z=qN -M\\. \label{spin1}
\end{equation}
The ground state and the low lying excitations of the model can be
studied, in the thermodynamic limit $N \rightarrow \infty$. In
this limit, the solutions of the Bethe ansatz equations are given
by the so called string hypothesis, i.e. the solutions of (\ref
{BAE1}) are grouped into strings of length $n$ with the same real
part and equidistant imaginary parts
\begin{equation}
\lambda_{\alpha}^{(n,j)}=\lambda_{\alpha}^n + {i\over
2}(n+1-2j),~~~j=1,2,...,n \label{STR}
\end{equation}
where $\lambda_{\alpha}^n$ is real.
 It is known \cite{KR}--\cite{fr1}, that the ground state, i.e. the state
with the least energy, is the filled Dirac sea with strings of
length $q=2s$. The low lying excitations are holes in the $q$ sea
and also strings of length $n \neq q$. One can study the
scattering among the low lying excitations of the model and show
\cite{KR}--\cite{BT} that the $S$ matrix, as $s \rightarrow
\infty$, coincides with the one of the $SU(2)$ $PCM$ i.e., it is
the $S_{SU(2)} \bigotimes S_{SU(2)}$ $S$ matrix \cite{ZAZ},
\cite{DB}. The $SU(2)$ scattering amplitudes (for the singlet
triplet and singlet respectively) are \be S_{0}( \lambda)= \exp
\Bigl \{-\int_{-\infty}^{\infty} e^{-i \omega \lambda}
{e^{-{\omega \over 2}} \over 2 \cosh({\omega \over 2})}{d\omega
\over \omega} \Bigr \},~~S_{+}( \lambda)= { \lambda + i \over
\lambda -i} S_{0}( \lambda). \label{S'+} \ee
 In the scaling limit, $\lambda << \Omega$,
 the excitations become massive relativistic particles \cite{fr1},
 with energy and momentum, \be \epsilon(\lambda)
\sim 2e^{-\pi \Omega} \cosh\pi \lambda, \qquad p(\lambda) \sim
2e^{-\pi \Omega} \sinh\pi \lambda \ee where the momentum $p$ is
defined $mod \pi$ for even number of excitations.

Faddeev and Reshetikhin showed in \cite{fr1} that for finite $s$,
inconsistencies in the counting of the states exist. Therefore,
the interpretation of the $S$ matrix as $S_{SG}(s) \bigotimes
S_{SU(2)}$ somehow fails; $S_{SG}(s)$ is the sine--Gordon $S$
matrix \cite{ZZ}, with the triplet amplitude being
\begin{eqnarray} S'_{0}(\lambda)=\exp \Bigl
\{-\int_{-\infty}^{\infty} e^{-i \omega \lambda} {\sinh\Bigl ((2s
-1){\omega \over 2}\Bigr ) \over 2 \cosh({\omega \over 2})
\sinh\Bigl (2s{\omega \over 2}\Bigr )}{d\omega \over \omega} \Bigr
\}\,,
\end{eqnarray}
and obviously $S_{SG}(s \to \infty) \rightarrow S_{SU(2)}$.
However, Reshetikhin conjectured in \cite{resh2} that the correct
$S$ matrix for the spin $s$ chain is the $S_{RSOS}(s) \bigotimes
S_{SU(2)}$ matrix which coincides with the $S$ matrix of the $WZW$
model at level $k=2s$ ($WZW_{k}$) \cite{ZAZ}, \cite{resh2},
\cite{DB}. The $S_{RSOS}(s)$ is the scattering matrix of the
$RSOS$ model, and the spin $s$ is related to the restriction
parameter $r$ of the $RSOS$ model i.e. $r=2s+2$. (for a more
detailed analysis see \cite{BR}, \cite{resh2}). More specifically,
the spin $s=1$ chain has a hidden super-symmetry, which
is described by the $RSOS$ part of the $S$ matrix \cite{resh2}. 

A key observation is that, as $k=2s \to \infty$, the $S_{RSOS}(s)
\bigotimes S_{SU(2)}$ matrix degenerates to a tensor product of
two rational matrices, and it coincides with the $S$ matrix of the
$SU(2)$ $PCM$ without topological term, found in \cite{PW},
\cite{fr1}, \cite{BT}. The later comment reflects the fact that
the perturbed $WZW_{k}$ as $k \to \infty$, reduces to the $PCM$
without topological term ($S_{RSOS}(s\to \infty)$ reduces to
$S_{SU(2)}$), as described in \cite{abl}.

  \subsection{The alternating $s^{1}$, $s^{2}$ spin chain}
Alternating spin chains have been originally introduced by de Vega
and
 Woyanorovich in \cite{VEWO} and they have been also studied in the bulk by several
 authors (see e.g. \cite{AM}--\cite{bado}). We define the transfer matrix of the chain with
inhomogeneities, \be t= tr_{0} T_{0}(\lambda) \ee where \be
T_{0}(\lambda) = R_{0 2N}^{1}(\lambda-\Omega)  R_{0
2N-1}^{2}(\lambda+\Omega)\cdots R_{0 2}^{1}(\lambda-\Omega)  R_{0
1 }^{2}(\lambda+\Omega) \,, \label{mono2} \ee and $R^{i}$ is
related to the spin $s^{i}$ ($i=1,2$) representation
(\ref{Rmatrix1}). The eigenavalues of the transfer matrix, after
we fuse the auxiliary space as well, are given by (see also
\cite{VEWO})
 \be && \Lambda_{l}^{(1,2)
}(\lambda) = \sum_{k=0}^{l} a_{k}^{(1)}(\lambda
-\Omega)^{N}a_{k}^{(2)}(\lambda +\Omega)^{N} \non\\
&& \prod_{j=1}^{M}\frac{(\lambda-\lambda_{j}-{i\over 2}
)}{(\lambda-\lambda_{j}+i(k-1)+{i\over2})}\frac{(\lambda-\lambda_{j}+
il +{i\over 2})}{(\lambda-\lambda_{j}+ik+{i\over2})},\ee where \be
a_{k}^{(j)}(\lambda) = \prod_{m=k}^{l-1}(\lambda -is^{j}+{i \over
2}+im)\prod_{n=0}^{k-1}(\lambda +is^{j} +in +{i\over 2}). \ee
  The corresponding Bethe ansatz
 equations have the
form, \be
e_{q^{1}}(\lambda_{\alpha}-\Omega)^{N}e_{q^{2}}(\lambda_{\alpha}+
\Omega)^{N}= -\prod_{\beta=1}^{M}
e_2(\lambda_{\alpha}-\lambda_{\beta}) \label{BEA2}, \ee  where
$q^{j} = 2s^{j}$ and $q^{1} - q^{2} =1$. The BEA  (\ref{BEA2})
coincide with the ones found by Polyakov and Wiegmann for the
lattice analogue of the $SU(2)$ $PCM$ with $WZW$ term \cite{PW1}.

Again for $l=2s^{j}$ ($j=1, 2$) and $\lambda= \pm \Omega +
\lambda_{0}^{j} = \pm \Omega -{i \over 2}(2s^{j} -1)$ the
$R^{j}(\lambda_{0}^{j})$ matrix becomes the permutation operator
and hence we obtain a local Hamiltonian, \be H \propto {i\over
2\pi} {d \over d\lambda} \log(t^{1}(\lambda))
\vert_{\lambda=\Omega +\lambda_{0}^{1}} +{i\over 2\pi}{d \over
d\lambda} \log(t^{2}(\lambda)) \vert_{\lambda=-\Omega
+\lambda_{0}^{2}}\ee and the corresponding eigenvalues are\be
E=-{1 \over 2 \pi }\sum_{n=1}^{2} \sum_{j=1}^{M} { q^{n} \over
(\lambda_{j}+(-)^{n} \Omega + {iq^{n}\over 2})(\lambda_{j}+(-)^{n}
\Omega- {iq^{n} \over 2})}. \label{energyc} \ee Moreover, the
momentum and spin of a Bethe state are given by \be P= {1 \over 2
i} \sum_{n=q^{1}}^{q^2} \sum_{j=1}^{M} \log {(\lambda _{j}+(-)^{n}
\Omega+ {iq^{n} \over 2}) \over( \lambda_{j}+(-)^{n} \Omega
-{iq^{n} \over 2})}\label{EPS2} \ee
\begin{equation}
S^z=(q^{1} + q^{2})N -M\\. \label{spin2}
\end{equation}

It has been proved \cite{DMN}, \cite{bado}, that this model has
two types of low lying excitations, i.e. holes. The scattering
among them was studied, the corresponding $S$ matrix was computed,
for $s^{i} \to \infty$, and it was shown to be the $S_{SU(2)}
\bigotimes S_{SU(2)}$ plus a non trivial left right scattering
$S_{LR}(\lambda)= \tanh{\pi \over 2}( \lambda -{i\over 2})$. The
$S_{SU(2)} \bigotimes S_{SU(2)}$, $S_{LR}(\lambda)$ is also the
massless $S$ matrix of the $SU(2)$ principal chiral model with
$WZW$ term at level 1 ($PCM_{1}$), conjectured by Zamolodchikovs
\cite{ZAZ}. It has been also proved \cite{bado}, that in the
scaling limit $\lambda << \Omega$, both excitations obey a
massless relativistic dispersion relation, namely
\begin{equation} \epsilon^{1}(\lambda)
= p^{1}(\lambda) \sim  e^{-\pi \Omega} e^{\pi \lambda}\,, \qquad
\epsilon^{2}(\lambda) = -p^{2}(\lambda) \sim e^{-\pi \Omega}
 e^{-\pi \lambda}\,. \label{em0} \end{equation}
 These are the energy
  and momentum of the ``right'' and ``left'' movers
 respectively  (see e.g. \cite{ZAZ}) and the factor
$e^{-\pi \Omega}$ provides a mass scale for the system.

\section{Spin chains with boundaries}
After the brief review on the bulk $XXX$ spin $s$ and alternating
chains, we are ready to study these models in the presence of
boundaries. For both models we will derive the Bethe ansatz
equations and we will compute the exact reflection matrices. To
construct the spin chain with boundaries in addition to the $R$
matrix another constructing element, the $K$ matrix, is needed.
The $K$ matrix is a solution of the reflection (boundary
Yang--Baxter) equation \cite{cherednik}, \be
R_{12}(\lambda_{1}-\lambda_{2})\ K_{1}(\lambda_{1})\
R_{2 1}(\lambda_{1}+\lambda_{2})\ K_{2}(\lambda_{2}) \non \\
= K_{2}(\lambda_{2})\  R_{12}(\lambda_{1}+\lambda_{2})\
K_{1}(\lambda_{1})\ R_{21}(\lambda_{1}-\lambda_{2}) \,.
\label{BYBE} \ee In what follows we are going to use Sklyanin's
formalism \cite{sklyanin} in order to construct both models with
boundaries.

\subsection{The $XXX$ spin $s$ open chain}

\subsubsection{The Bethe ansatz equations}

Let as first consider the $XXX$ spin $s={q\over 2}$ quantum chain
with boundaries. We define the transfer matrix for the open chain
\cite{sklyanin}
 \be t(\lambda) = \tr_{0}
K_{0}^{+}(\lambda)\ T_{0}(\lambda)\ K^{-}_{0}(\lambda)\ \hat
T_{0}(\lambda)\,, \label{transfer1} \ee  $T_{0}(\lambda)$ is the
monodromy matrix (\ref{monodromy1}) and \be \hat T_{ 0}(\lambda) =
R_{1 0}^{2s}(\lambda-\Omega) R_{2 0}^{2s}(\lambda+\Omega)\cdots
R_{2N-1 0}^{2s}(\lambda-\Omega) R_{2N 0}^{2s}(\lambda+\Omega) \,.
\label{monodromy2} \ee
 The $K^{-}(\lambda)=K(\lambda,\xi^{-})$
matrix satisfies the reflection equation (\ref{BYBE}) and the
$K^{+}(\lambda)=K^{-}(-\lambda-\rho)^{t}$ ($\xi^{-} \to -\xi^{+}$)
satisfies,
 \be R_{12}(-\lambda_{1}+\lambda_{2})\
K_{1}^{+}(\lambda_{1})^{t_{1}}\
R_{2 1}(-\lambda_{1}-\lambda_{2}-2\rho)\ K_{2}^{+}(\lambda_{2})^{t_{2}} \non \\
= K_{2}^{+}(\lambda_{2})^{t_{2}}\
R_{12}(-\lambda_{1}-\lambda_{2}-2\rho)\
 K_{1}^{+}(\lambda_{1})^{t_{1}}\ R_{21}(-\lambda_{1}+\lambda_{2}) \,,
\label{boundaryYB2} \ee where $\rho = i$. We choose the diagonal
solution of the reflection equation \cite{GZ}, \cite{DVGR3},
namely \be K(\lambda) = diag (-\lambda +i\xi, ~~~ \lambda +i\xi).
\label{gz} \ee
 Note that
in the presence of boundaries we have to fuse the $K$ matrix as
well, and we use the following fusion hierarhy for the transfer
matrix \cite{mn/fusion}, \cite{nxxz} \be t^{j}(\lambda) &=& \tilde
\zeta_{2j-1}(2\lambda +(2j-1)i)\Bigl
[ t^{j-{1 \over 2}}(\lambda)t^{1 \over 2}(\lambda+(2j-1)i) \non\\
&-& {\Delta(\lambda +(2j-2)i) \tilde \zeta_{2j-2}(2\lambda
+(2j-2)i) \over \zeta(2\lambda +2i(2j-1))} t^{j-1}(\lambda) \Bigr
] \label{fusion} \ee where \be \Delta(\lambda) =
\Delta[K^{+}(\lambda)]\ \delta[T(\lambda)]\
\Delta[K^{-}(\lambda)]\ \delta[\hat T(\lambda)]\,, \ee and the
quantum determinants are \cite{sklyanin}, \cite{KR},
\cite{mn/fusion}, \be & & \delta[T(\lambda)] = \delta[\hat
T(\lambda)] = \hat \zeta(\lambda + \Omega+i)^{N} \hat
\zeta(\lambda -\Omega+i)^{N}, \non\\ & & \Delta[K^{-}(\lambda)] =
g(2\lambda + \rho)l(\lambda, \xi^{-}), \qquad
\Delta[K^{+}(\lambda)] = g(-2\lambda -3\rho)l(\lambda, \xi^{+})
\label{det} \ee moreover \be \zeta(\lambda) &=& (\lambda
+i)(-\lambda+i),~~\hat \zeta(\lambda) = (\lambda +is +{i\over
2})(-\lambda+is+{i\over 2}), \non\\
g(\lambda) &=& -\lambda+i, ~~l(\lambda, \xi)= (\lambda+ i
+\xi)(\lambda+ i -\xi) \ee and \be \tilde \zeta_{j}(\lambda) =
\prod_{k=1}^{j}\zeta(\lambda + ik), ~~\tilde \zeta_{0}(\lambda)=1.
\ee
 Then the eigenvalues of the transfer matrix, after we fuse the auxiliary space,
  are given by
 \be \Lambda_{l}^{2s}(\lambda) &=&
\sum_{k=0}^{l}h_{k}(\lambda) f_{k}(\lambda) a_{k}(\lambda
-\Omega)^{2N}a_{k}(\lambda +\Omega)^{2N}\non\\ &&
\prod_{j=1}^{M}\frac{(\lambda-\lambda_{j}-{i\over 2}
)}{(\lambda-\lambda_{j}+i(k-1)+{i\over2})}\frac{(\lambda-\lambda_{j}+
il +{i\over 2})}{(\lambda-\lambda_{j}+ik+{i\over2})} \non\\
&& \prod_{j=1}^{M}\frac{(\lambda+\lambda_{j}-{i\over 2}
)}{(\lambda+\lambda_{j}+i(k-1)+{i\over2})}\frac{(\lambda+\lambda_{j}+
il +{i\over 2})}{(\lambda+\lambda_{j}+ik+{i\over2})} \ee where
$a_{k}$ have been already defined in (\ref{a_{k}}) and  \be
f_{k}(\lambda) = f_{k}^{+}(\lambda)f_{k}^{-}(\lambda) \ee where
\be f_{k}^{\pm}(\lambda)=\prod_{m=k}^{l-1}(\lambda +i \xi^{\pm} +
im+i) \prod_{n=0}^{k-1} (-\lambda +i \xi^{\pm} -in) \ee
$f_{k}^{\pm}(\lambda)$ correspond to the left (+) and right (-)
boundary. $h_{k}(\lambda)$ are derived by the fusion hierarhy
(\ref{fusion}) and for e.g. $l=1$, $l=2$  they are respectively
\footnote{For general $l$ \be h_{k}(\lambda) \propto
\prod_{m=k}^{l-1}(2\lambda + 2im) \prod_{n=0}^{k-1} (2\lambda +2i+
2in). \ee} \be h_{k}(\lambda) ={1 \over 2 \lambda
+i}\prod_{m=k}^{l-1}(2\lambda + 2im) \prod_{n=0}^{k-1} (2\lambda
+2i+ 2in) \ee and \be h_{k}(\lambda) =\prod_{m=k}^{l-1}(2\lambda +
2im) \prod_{n=0}^{k-1} (2\lambda +2i+ 2in). \ee
 From the
analyticity of the eigenvalues we obtain the Bethe ansatz
equations \be &&
e_{x^{+}}^{-1}(\lambda_{\alpha})e_{x^{-}}^{-1}(\lambda_{\alpha})
e_{q}(\lambda_{\alpha}-\Omega)^{2N}e_{q}(\lambda_{\alpha}+\Omega)^{2N}e_{1}(\lambda_{\alpha})
\non\\ && = -\prod_{\beta=1}^{M}
e_2(\lambda_{\alpha}-\lambda_{\beta})e_2(\lambda_{\alpha}+\lambda_{\beta})\,,
\label{BE}\ee where $x^{\pm} = 2 \xi^{\pm}+1$. Recall that for
$l=2s$, $R^{2s}(\lambda_{0})$ is the permutation operator,
therefore we obtain a local Hamiltonian
 \be H \propto {i\over 2 \pi}{d \over d\lambda} \log(t^{2s}(\lambda))
\vert_{\lambda=\pm \Omega +\lambda_{0}} \ee and the corresponding
energy eigenvalues \be E=-{1 \over 2 \pi
}\sum_{n=1}^{2}\sum_{j=1}^{M} \ { q \over (\lambda_{j}+(-)^{n}
\Omega + {iq \over 2})(\lambda_{j}+(-)^{n} \Omega- {iq \over 2})}.
\label{energyd} \ee

\subsubsection{The boundary $S$ matrix}

Now we can study the scattering of the excitations with the
boundaries and determine the boundary $S$ matrix. Before we do
that it is necessary to determine what is the ground state and the
low lying excitations for the model.
 The ground state, as in the bulk, is the filled Dirac sea with strings of
length $q=2s$. The Bethe ansatz equations for the ground state
become \be &&
X_{qx^{+}}^{-1}(\lambda_{\alpha})X_{qx^{-}}^{-1}(\lambda_{\alpha})X_{qq}(\lambda_{\alpha}-\Omega)^{2N}
X_{qq}(\lambda_{\alpha}+\Omega)^{2N}e_{q}(\lambda_{\alpha}) \non\\
&& =
 \prod_{\beta=1}^{M_{j}} E_{qq}(\lambda_{\alpha}-\lambda_{\beta})
 E_{qq}(\lambda_{\alpha}+\lambda_{\beta}).
\label{K1} \ee Let us consider $x^{\pm} \geq q-1$, and define
\begin{eqnarray}
X_{nm}(\lambda)&=& e_{|n-m+1|}(\lambda) e_{|n-m+3|}(\lambda)\ldots
e_{(n+m-3)}(\lambda)e_{(n+m-1)}(\lambda)
\nonumber\\
E_{nm}(\lambda)&=& e_{|n-m|}(\lambda)e_{|n-m+2|}^{2}(\lambda)
\ldots e_{(n+m-2)}^{2}(\lambda)e_{(n+m)}(\lambda).\end{eqnarray}
For the next we need the following notations
\begin{eqnarray}
a_{n}(\lambda) = {1\over 2 \pi} {d \over d\lambda} i \log
e_{n}(\lambda) \ee and the Fourier transform of $a_{n}$, is given
by
\begin{equation}
\hat a_{n}(\omega) = e^{ -{n\omega \over 2}}.
\end{equation}
We also need the following expressions
\begin{equation}
\Bigl (Z_{nm}(\lambda),\ A_{nm}(\lambda)\Bigr )= {1 \over 2\pi}{d
\over d\lambda} i \log \Bigl (X_{nm}(\lambda),\
E_{nm}(\lambda)\Bigr )\,,
\end{equation}
where their Fourier transforms are
\begin{equation}
\hat Z_{nm}(\omega)= { e^{ - {\max(n,m)\omega \over 2}} \sinh
\Bigl ((\min(n,m)){\omega \over 2} \Bigr ) \over  \sinh({\omega
\over 2})}\\,
\end{equation}
\begin{equation}
\hat A_{nm}(\omega)=2 \coth ({\omega \over 2} ) e^{ -
{\max(n,m)\omega \over 2}} \sinh \Bigl ((\min(n,m)){\omega \over
2}\Bigr )-\delta_{nm}\\.
\end{equation}

Finally, the energy (\ref{energyd}) takes the form
\begin{equation}
E = -  \sum_{n=1}^{2}\sum_{\alpha =1}^{M}Z_{qq}
(\lambda_{\alpha}+(-)^{n}\Omega)\\. \label{energyp1}
\end{equation}
The density for the ground state is given by \be
\sigma_{0}(\lambda)&=& Z_{qq}(\lambda+\Omega) +
Z_{qq}(\lambda-\Omega) -(A_{qq}*\sigma_{0})(\lambda) \non\\ &
+&{1\over L}\Bigl
(-Z_{qx^{-}}(\lambda)-Z_{qx^{+}}(\lambda)+a_{q}(\lambda)+A_{qq}(\lambda)\Bigr
) \ee where $L=2N$ is the length of the chain and $*$ denotes the
convolution. The solution of the last equation is \be
\sigma_{0}(\lambda)= \epsilon(\lambda) +{1\over L} \Bigl (
r(\lambda, \xi^{+})+ r(\lambda, \xi^{-})+ Q(\lambda)\Bigr ) \ee
where \be  \hat Q(\omega) = {\hat a_{q}(\omega) \over 1 +\hat
A_{qq} (\omega)} + \hat K_{1}(\omega), ~~\hat r(\omega,\xi^{\pm})
=-{ \hat Z_{qx^{\pm}}(\omega) \over 1+ \hat A_{qq}(\omega)},
~~\hat K_{1}(\omega) = {\hat A_{qq}(\omega) \over 1+ \hat
A_{qq}(\omega)}, \label{K_{1}} \ee and
\begin{equation} \epsilon(\lambda) = \sum_{i=1}^{2}
s(\lambda-(-)^{i} \Omega),~~ \hat s(\omega)=  {\hat Z_{qq}(\omega)
\over 1+\hat A_{qq}(\omega)}.
\end{equation} We can write the above Fourier transforms in terms
of trigonometric functions using the definitions of $\hat a_{q}$
and $\hat A_{qq}$, \be \hat K_{1}(\omega) = { e^{{(q-1) \omega
\over 2}} - e^{-{q \omega \over 2}}\cosh({\omega\over 2}) \over 2
\cosh({\omega \over 2})\sinh({q\omega \over 2})},~~\hat Q(\omega)
=\hat K_{1}(\omega) + { 1 \over 2 \coth({\omega \over
2})\sinh({q\omega \over 2})} \ee and for $x^{\pm} \geq q-1$ \be
\hat r(\omega, \xi^{\pm}) = -{e^{-(x^{\pm}-q){\omega \over 2}}
\over 2 \cosh({\omega \over 2})} \ee moreover,
\begin{equation}
\hat s(\omega)=  {1\over2 \cosh({\omega \over 2})}, ~~
\epsilon(\lambda) =\sum_{i=1}^{2}{1 \over 2 \cosh\Bigl
(\pi(\lambda-(-)^{i} \Omega)\Bigr)}\,. \label{energy2}
\end{equation}

 Let us consider the
state with $\nu$ holes in the $q$ sea, where $\nu$ an even number.
Then, in the thermodynamic limit we obtain the density of the
state form the Bethe ansatz equation, namely \be \sigma(\lambda)
&=& \sigma_{0}(\lambda)+ {1\over L}\sum_{i=1}^{\nu}
\Bigl(K_{1}(\lambda-\lambda_{i})+K_{1}(\lambda+\lambda_{i})\Bigr).
\label{densitya}\ee The energy of the state with $\nu$ holes in
the $q$ sea (\ref{energyp1}) is given
 by
\begin{equation}
E= E_{0} +  \sum_{\alpha=1}^{\nu} \epsilon(\lambda_{\alpha})\\,
\end{equation}
where $E_{0}$ is the energy of the ground state and
$\epsilon(\lambda)$ is the energy of the hole in the $q$ sea.
 Finally, we
compute the spin of the holes, and we can see from (\ref{spin1})
that the spin of a hole in the $q$ sea is (\ref{spin1}) $s={f
\over 2}$, where $f$ is an overall factor (see also \cite{KR}). We
consider the spin of the hole to be ${1\over 2}$ for what follows.
We conclude that the hole in the $q$ sea is a particle like
excitation with energy $\epsilon$, momentum $p$
($\epsilon(\lambda) = {1\over \pi} {d \over d\lambda}p(\lambda)$)
\be \epsilon(\lambda) = \sum_{n=1}^{2}{1 \over 2 \cosh (\pi
\lambda -(-)^{n} \Omega)}, ~~p(\lambda)= \pm {\pi \over 4} +
\sum_{n=1}^{2}{1\over 2} \tan^{-1}\Bigl (\sinh (\pi \lambda
-(-)^{n} \Omega)\Bigr ) \label{momentum1} \ee and spin $s={1 \over
2}$. We can easily check that in the scaling limit where $\lambda
<< \Omega $ the energy and momentum become from (\ref{momentum1}),
\be \epsilon(\lambda) \sim 2e^{- \pi \Omega} \cosh(\pi \lambda),
~~p(\lambda) \sim 2e^{-\pi \Omega}
 \sinh(\pi \lambda)\,. \ee
Note that the excitations in the scaling limit, as in the bulk
case satisfy a massive relativistic dispersion relation.

 Having studied the excitations
of the model we are ready to compute the complete boundary
$S$-matrix. To do so we follow the formulation developed by
Korepin, and later by Andrei and Destri \cite{K}, \cite{AD}. First
we have to consider the so called quantization condition.
\begin{equation}
(e^{2iLp}S-1)|\tilde \lambda_{1}, \tilde \lambda_{2} \rangle = 0
\label{qc1}
\end{equation}
where
 $p$ is the momentum of
the particle, the hole in our case. For the case of $\nu$ (even)
holes in the $q$ sea, we compare the integrated density
(\ref{densitya}) with the quantization condition (\ref{qc1}).
Having also in mind that,
\begin{equation}
\epsilon(\lambda) = {1 \over  \pi} {d \over d \lambda} \
p(\lambda)
\end{equation}
we end up with the following expression for the boundary
scattering amplitudes (the boundary $S$ matrix will have the form
$S^{\pm}=diag(\alpha^{\pm}, \beta^{\pm})$), \be \alpha^{+}
\alpha^{-} = \exp \Bigl \{ 2piL \int_{0}^{\tilde
\lambda_{1}}d\lambda \Bigl (\sigma(\lambda)
-\epsilon(\lambda)\Bigr) \Bigr \} \ee and
 \be \alpha^{\pm}(\lambda, \xi^{\pm}) = f(
\lambda) k_{0}( \lambda)k_{0}( \lambda) k_{1}(\lambda, \xi^{\pm})
\label{sol1} \ee where
\begin{equation}
k_{0}(\tilde \lambda_{1})= \exp \Bigl \{2 \pi i \int_{0}^{\tilde
\lambda_{1}}{1\over 4}\Bigl (\sum_{i=1}^{\nu}(K_{1}(\lambda-
\tilde \lambda_{i})+K_{1}(\lambda+ \tilde \lambda_{i}))
+Q(\lambda)\Bigr )
 d \lambda \Bigr \} \label{sol3}
\end{equation} the $x$ dependent part is
\begin{equation}
k_{1}(\tilde \lambda_{1}, \xi^{\pm})= \exp \Bigl \{2 \pi i
\int_{0}^{\tilde \lambda_{1}} r(\lambda, \xi^{\pm})  d \lambda
\Bigr \}. \label{sol4}
\end{equation}
We are interested in the limit that $s \rightarrow \infty$, in
this limit we can easily see from (\ref{K_{1}}) (see also
\cite{W1}) that \be \hat Q(\omega) \rightarrow \ \hat
K_{1}(\omega),
 ~~\hat K_{1}(\omega)  \rightarrow  {e^{-{\omega \over 2}} \over
\cosh{\omega \over 2}}, ~~ \hat r(\omega, \xi^{\pm}) =
-{e^{-(x^{\pm}-q){\omega \over 2}} \over 2 \cosh({\omega \over
2})} \ee with $x^{\pm}-q$ to be a fixed number as $q \rightarrow
\infty$ and $f(\lambda)$ is an overall CDD factor given by
\begin{equation}
f( \lambda)= \exp \Bigl \{\int_{-\infty}^{\infty} e^{-i \omega
 \lambda} {1 \over 2 \cosh({\omega \over 2})}{d\omega \over
\omega} \Bigr \}=\tanh{\pi \over 2}(\lambda+{i\over 2})\\.
\end{equation}
 Using the above Fourier transforms we end up with the following
expressions,   \be k_{0}(\lambda) &=& \exp \Bigl
\{2\int_{0}^{\infty} {d\omega \over \omega} \sinh (2i\omega
\lambda){\sinh({3\omega \over 2}) e^{-{\omega \over 2}} \over
\sinh(2\omega)} \Bigr \},\non\\ k_{1}(\lambda, \tilde \xi^{\pm})
&=& \exp \Bigl \{-\int_{0}^{\infty} {d\omega \over
\omega}\sinh(2i\omega \lambda){ e^{-2 \tilde \xi^{\pm} \omega}
\over \cosh(\omega)}\Bigr \}, \label{sol5} \ee  where $2\tilde
\xi^{\pm} =2 \xi^{\pm}-2s+1$ is the renormalized boundary
parameter and $k_{0}(\lambda)$, $k_{1}(\lambda, \xi^{\pm})$ are
the $\xi$ independent and the $\xi$ dependent part, respectively
of the usual $XXX$ ($SU(2)$) reflection matrix (see e.g.
\cite{gmn}). We notice from (\ref{sol1}) that there are two copies
of the $\xi$ independent part, whereas just one copy of the $\xi$
dependent part exists.
 Recall, that we considered the diagonal $K$ matrix (\ref{gz}) in order to
 construct our chain, therefore we need to determine the other element of
 the boundary matrix. To do so we exploit the ``duality'' symmetry
\cite{done} of the transfer matrix for $\xi^{\pm} \rightarrow
-\xi^{\pm}$. Hence, the other element is given by \be
\beta^{\pm}(\lambda, \tilde \xi^{\pm}) = e_{2 \tilde
\xi^{\pm}-1}(\lambda) \alpha^{\pm}(\lambda, \tilde \xi^{\pm}),
\label{sol0} \ee the ratio ${\beta^{\pm}(\lambda, \tilde
\xi^{\pm}) \over \alpha^{\pm}(\lambda, \tilde \xi^{\pm})}$ is the
same as in the $XXX$ model \cite{gmn} but with a renormalized
boundary parameter.

The above equations (\ref{sol1})--(\ref{sol0}), are simply
combined to give the
 boundary $S$ matrix as the tensor product of two rational ($XXX$)
boundary $S$ matrices \cite{gmn} (up to an overall CDD  factor)
i.e. \be S(\lambda) = f(\lambda) S_{SU(2)}(\lambda, \tilde
\xi^{\pm}_{1} ) \bigotimes S_{SU(2)}(\lambda, \tilde
\xi^{\pm}_{2}), \label{sm}\ee
 where $\tilde
\xi^{\pm}_{1}=\tilde \xi^{\pm}$, $\tilde \xi^{\pm}_{2} \to
\infty$. This matrix coincides with the one found for the $SU(2)$
$PCM$ \cite{mash}
---in our case $\tilde \xi^{\pm}$ is a free parameter.

However, in analogy with the bulk case, for finite $s$ the
boundary $S$ matrix is expected to be of the form
$S_{SU(2)}(\lambda, \tilde \xi^{\pm}) \bigotimes S_{RSOS}(s)$,
where $S_{RSOS}(s)$ is the boundary $S$ matrix of the $RSOS$
model. The $S_{RSOS}$ matrix is a solution of the boundary
Yang--Baxter equation in the $RSOS$ representation (see e.g.
\cite{rsosb1}--\cite{rsosb3}). The $S_{SU(2)}(\lambda, \xi^{\pm} )
\bigotimes S_{RSOS}(s)$ matrix should also describe the
$WZW_{k=2s}$ boundary scattering and it should presumably reduce
to the $S$ matrix we found (\ref{sm}), as $s \to \infty$.

\subsection{The alternating $s^{1}$, $s^{2}$ open spin chain}

\subsubsection{The Bethe ansatz equations}

 The corresponding transfer matrix $t(\lambda)$ for
the open alternating chain of $2N$ sites and $s^{1}={q^{1}\over
2}$, $s^{2}={q^{2}\over 2}$ spins $(q^{1}-q^{2}=1)$ is (see also
e.g., \cite{VEWO}, \cite{sklyanin}),  \be t(\lambda) = \tr_{0}
K_{0}^{+}(\lambda)\ T_{0}(\lambda)\ K^{-}_{0}(\lambda)\ \hat
T_{0}(\lambda)\,, \label{transfer2} \ee where $T_{0}(\lambda)$ is
the monodromy matrix defined previously in (\ref{mono2}) and \be
\hat T_{ 0}(\lambda) = R_{1 0}^{2}(\lambda-\Omega) R_{2
0}^{1}(\lambda+\Omega)\cdots R_{2N-1 0}^{2}(\lambda-\Omega) R_{2N
0}^{1}(\lambda+\Omega) \,. \label{hatmonodromy} \ee Again here we
use the diagonal $K$ matrix (\ref{gz}) and the fusion hierarchy
(\ref{fusion}) with \be \delta[T(\lambda)] &=& \hat
\zeta_{1}(\lambda - \Omega +i)^{N} \hat \zeta_{2}(\lambda +\Omega
+i)^{N}, \non\\ \delta[\hat T(\lambda)]&=& \hat \zeta_{1}(\lambda
+ \Omega +i)^{N} \hat \zeta_{2}(\lambda -\Omega+i)^{N} \ee and \be
\hat \zeta_{n}(\lambda) = (-\lambda+is^{n}+{i\over
2})(\lambda+is^{n}+{i\over 2}).\ee
 The eigenvalues of the transfer matrix,
after we fuse the auxiliary space, are given by \be
\Lambda_{l}^{(1,2)}(\lambda) &=& \sum_{k=0}^{l}h_{k}(\lambda)
f_{k}(\lambda) a_{k}^{(1)}(\lambda -\Omega)^{N}a_{k}^{(1)}(\lambda
+\Omega)^{N}a_{k}^{(2)}(\lambda -\Omega)^{N}a_{k}^{(2)}(\lambda
+\Omega)^{N} \non\\ &&
\prod_{j=1}^{M}\frac{(\lambda-\lambda_{j}-{i\over 2}
)}{(\lambda-\lambda_{j}+i(k-1)+{i\over2})}
\frac{(\lambda-\lambda_{j}+ il +{i\over 2})}
{(\lambda-\lambda_{j}+ik+{i\over2})} \non\\ &&
\prod_{j=1}^{M}\frac{(\lambda+\lambda_{j}-{i\over 2}
)}{(\lambda+\lambda_{j}+i(k-1)+{i\over2})}\frac{(\lambda+\lambda_{j}+
il +{i\over 2})}{(\lambda+\lambda_{j}+ik+{i\over2})} \ee
 provided that $\{ \lambda_{j} \}$ satisfy the Bethe ansatz
equations \be && \prod_{n=q^{1}}^{q^{2}}
e_{n}(\lambda_{\alpha}-\Omega)^{N}e_{n}(\lambda_{\alpha}+\Omega)^{N}
e_{1}(\lambda_{\alpha}) e_{x^{+}}^{-1}
(\lambda_{\alpha})e_{x^{-}}^{-1} (\lambda_{\alpha}) \non\\ &&=
-\prod_{\beta=1}^{M}
e_{2}(\lambda_{\alpha}-\lambda_{\beta})e_{2}(\lambda_{\alpha}+\lambda_{\beta}).
\label{BE2} \ee  For $l =2s^{j}$, $R^{j}(\lambda_{0}^{j})$ becomes
the permutation operator, therefore we can obtain a local
Hamiltonian for the open chain
 \be H \propto {i\over 4 \pi}{d \over d\lambda}
\log(t^{1}(\lambda)) \vert_{\lambda=\Omega +\lambda_{0}^{1}}
+{i\over 4 \pi}{d \over d\lambda} \log(t^{2}(\lambda))
\vert_{\lambda=-\Omega +\lambda_{0}^{2}}\ee and \be E&=&-{1 \over
4 \pi }\sum_{n=1}^{2} \sum_{j=1}^{M} \ { q^{n} \over (\lambda_{j}-
\Omega + {iq^{n} \over 2})(\lambda_{j}- \Omega- {iq^{n} \over 2})}
\non\\ &-&{1 \over 4 \pi }\sum_{n=1}^{2}\sum_{j=1}^{M} \ { q^{n}
\over (\lambda_{j}+ \Omega + {iq^{n} \over 2})(\lambda_{j}+
\Omega- {iq^{n} \over 2})}. \label{energyf} \ee

\subsubsection{The boundary $S$ matrix}

The ground state consists of two filled Dirac seas with strings of
length $q^{n}=2s^{n}$, $n=1,2$. Then the Bethe ansatz equations
for the ground state become \be &&
X_{nx^{+}}^{-1}(\lambda_{\alpha}^n)X_{nx^{-}}^{-1}(\lambda_{\alpha}^n)\prod
_{j=q^{1}}^{q^{2}}X_{nj}(\lambda_{\alpha}^n-\Omega)^{N}X_{nj}(\lambda_{\alpha}^n+\Omega)^{N}
e_{n}(\lambda_{\alpha}^{n})\non\\
&&= \prod_{j= q^{1}}^{q^{2}} \prod_{\beta=1}^{M_{j}}
E_{nj}(\lambda_{\alpha}^{n}-\lambda_{\beta}^j)E_{nj}(\lambda_{\alpha}^{n}+\lambda_{\beta}^j)
\label{K2} \ee where $n$ can be $q^{1}$, $q^{2}$ and let us
consider for simplicity $x^{\pm} \geq q^{2}$.
Finally, the energy (\ref{energyf}) takes the form
\begin{equation}
E = -{1\over 2} \sum_{i,j= q^{1}}^{q^{2}} \sum_{\alpha
=1}^{M_{j}}Z_{ij} (\lambda_{\alpha}^{i}+\Omega) - {1\over
2}\sum_{i,j= q^{1}}^{q^{2}} \sum_{\alpha =1}^{M_{j}}Z_{ij}
(\lambda_{\alpha}^{i}-\Omega)\\. \label{energyp2}
\end{equation}
The density that describes the ground state is given by the
following integral equations, \be \sigma_{0}^{n}(\lambda) &=&
\sum_{j= q^{1}}^{q^{2}} {1\over 2}
\Bigl(Z_{nj}(\lambda-\Omega)+Z_{nj}(\lambda+ \Omega) \Bigr ) -
\sum_{ j= q^{1}}^{q^{2}} (A_{nj}*\sigma_{0}^{j})(\lambda) \non\\
&+& {1\over L}\Bigl (-Z_{nx^{+}}(\lambda) -Z_{nx^{-}}(\lambda)+
a_{n}(\lambda)+ \sum_{j= q^{1}}^{q^{2}}A_{nj}(\lambda)\Bigr )
\label{density1} \ee where $n$ can be $q^{1}$ or $q^{2}$. The
solution of the above integral equation is given by \be
\sigma_{0}^{n}(\lambda) = \epsilon^{n}(\lambda)+{1\over
L}\Bigl(r^{n}(\lambda, \xi^{+})+ r^{n}(\lambda, \xi^{-})+
Q^{n}(\lambda)\Bigr) \ee where \be \hat Q^{n}(\omega) &=&
\sum_{i=q^{1}}^{q^{2}} \hat a_{i}(\omega)\hat R_{ni}(\omega)+
\sum_{i=q^{1}}^{q^{2}}\hat K_{1}^{in}(\omega), \non\\
\hat r^{n}(\omega,\xi^{\pm}) &=& -\sum_{i=q^{1}}^{q^{2}} \hat
Z_{ix^{\pm}}(\omega) \hat R_{ni}(\omega), ~~~\hat
K_{1}^{nm}(\omega) = \sum_{i=q^{1}}^{q^{2}}\hat A_{ni}(\omega)\hat
R_{mi}(\omega) \ee and
\begin{equation}
\epsilon^{n} (\lambda) = {1\over 2}\sum_{i=1}^{2}
s(\lambda-(-)^{i} \Omega)\,, \label{energy2b}
\end{equation}
and its Fourier transform is
\begin{equation}
\hat s^{n}(\omega)= \sum_{i, j=q^{1}}^{q^{2}} (\hat Z_{ij} \hat
R_{nj})(\omega)\\. \label{energy02}
\end{equation}
Here $R$ is the inverse of the kernel $K$ of the system of the
linear equations (\ref{density1}),
\begin{equation}
\hat K_{nm}(\omega)=(1 +\hat A_{nm}(\omega)) \delta_{nm} + \hat
A_{nm}(\omega)(1-\delta_{nm})
\end{equation}
\begin{equation}
\hat R_{nm}(\omega)= {1 \over det \hat K}
\sum_{j=q^{1}}^{q^{2}}((1 +\hat A_{jj}(\omega))
\delta_{nm}(1-\delta_{nj}) - \hat
A_{nm}(\omega)(1-\delta_{nm}))\\,
\end{equation}
where the determinant of $K$ is, in terms of trigonometric
functions,
\begin{equation}
det \hat K = 4 \coth^{2} ({\omega \over 2}) e^{- {q^{1}\omega
\over 2}} \sinh (q^{2}{\omega \over 2}) \sinh ({\omega \over
2})\\.
\end{equation}
In particular, $\hat K_{1}^{nm}$ has the following explicit form
in terms of trigonometric functions \be \hat K_{1}^{11}(\omega)&=&
{e^{-{\omega \over 2}} \over 2 \cosh({\omega \over 2})},~~ \hat
K_{1}^{22}(\omega)= {\sinh \Bigl ((q^{1} -2){\omega \over 2}\Bigr
) \over 2\cosh({\omega \over 2}) \sinh \Bigl ((q^{1} -1){\omega
\over 2}\Bigr )} \non\\ \hat K_{1}^{12}(\omega)&=& \hat
K_{1}^{21}(\omega)={1 \over 2 \cosh({\omega \over 2})} \label{K11}
\ee \be \hat Q^{1}(\omega) = {1 +e^{-{\omega \over 2}} \over 2
\cosh({\omega \over 2})}, ~~\hat Q^{2}(\omega) = \hat
K^{22}(\omega)+ {1\over 2 \cosh({\omega \over 2})} +
{\sinh({\omega \over 2}) \over 2 \cosh({\omega \over 2})
\sinh({q^{2} \omega \over 2})}. \ee The $\xi$ dependent part for
$x^{\pm}\geq q^{2}$ is
 \be \hat r^{1}(\omega, \xi^{\pm})
=-{e^{-(x^{\pm} -q^{1}){\omega \over 2}} \over 2 \cosh({\omega
\over 2})}, ~~\hat r^{2}(\omega,\xi^{\pm}) =0 \ee
finally,
\begin{equation}
\hat s^{n}(\omega)= {1\over 2 \cosh({\omega \over
2})},~~\epsilon^{n} (\lambda) =\sum_{i=1}^{2}{1 \over 4 \cosh\Bigl
(\pi(\lambda+(-)^{i} \Omega)\Bigr)}\,.\label{energy2a}
\end{equation}

Let us consider the state with $\nu_{n}$ holes in the $q^{n}$ sea,
where $\nu_{n}$ is an even number. Then in the thermodynamic limit
we obtain the density of the state from the Bethe ansatz
equations, namely \be \sigma^{n}(\lambda) =
\sigma_{0}^{n}(\lambda) + {1\over L}\sum_{i=1}^{\nu_{n}}
\Bigl(K_{1}^{nn}(\lambda-\lambda_{i})+K_{1}^{nn}(\lambda+\lambda_{i})\Bigr)
\label{densityb} \ee the energy of the state with $\nu_{n}$ holes
in the $q^{n}$ seas is given (\ref{energyp2}) by
\begin{equation}
E= E_{0} +  \sum_{\alpha=1}^{\nu_{n}}
\epsilon^{n}(\lambda_{\alpha}^{n})\\,
\end{equation}
where $E_{0}$ is the energy of the ground state and
$\epsilon^{n}(\lambda)$ is the energy of the hole in the $q^{n}$
sea.
 Finally, we
compute the spin of the holes from (\ref{spin2}), and we can
verify that the spin of a hole in the $q^{1}$ sea is $s^{1}={1
\over 2}$ whereas the spin of a hole in the $q^{2}$ sea is
$s^{2}=0$. We conclude that the hole in the $q^{n}$ sea is a
particle like excitation with energy $\epsilon^{n}$, momentum
$p^{n}$ ($\epsilon^{n}(\lambda) = {1\over \pi} {d \over
d\lambda}p^{n}(\lambda)$) \be \epsilon^{n}(\lambda) &=&
\sum_{l=1}^{2}{1 \over 4 \cosh\Bigl (\pi(\lambda-(-)^{l}
\Omega)\Bigr)}, \non\\ p^{n}(\lambda)&=&\pm {\pi \over 4} +
\sum_{l=1}^{2}{1\over 4} \tan^{-1}\Bigl (\sinh\pi (\lambda-(-)^{l}
\Omega)\Bigr ), \label{momentum2} \ee  and spin $s^{1}={1 \over
2}$, $s^{2} = 0$. We can easily check that in the scaling limit,
$\lambda << \Omega$ the energy and momentum become from
(\ref{momentum2}), \be \epsilon^{n}(\lambda) \sim e^{-\pi \Omega}
\cosh(\pi \lambda)\,, ~~ p^{n}(\lambda) \sim e^{-\pi \Omega}
 \sinh(\pi \lambda)\,, \label{em} \ee
 the factor
$e^{-\pi \Omega}$ provides a mass scale for the system. Note that
in the presence of boundaries the excitations, in the scaling
limit, satisfy a massive relativistic dispersion relation
(\ref{em}) whereas in the bulk case the excitations are massless
relativistic particles (\ref{em0}). This is a very interesting
phenomenon which is presumable related to the type of the
boundaries. The boundaries we impose, force a left (right) mover
to reflect as a left (right) mover. It is possible that the
boundary can reflect a left mover to a right one and vice versa.
This type of boundaries would probably lead to massless
excitations in the scaling limit.

Again, we consider the quantization condition \cite{K}, \cite{AD},
in order to compute the exact reflection matrices, namely
\begin{equation}
(e^{2iLp^{n}}S^{n}-1)|\tilde \lambda_{1}, \tilde \lambda_{2}
\rangle = 0 \label{qc2}
\end{equation}
where $p^{n}$ is the momentum of the hole in the $q^{n}$ sea. For
the case of $\nu^{n}$ holes in the $q^{n}$ sea, we compare the
integrated density (\ref{densityb}) with the quantization
condition (\ref{qc2}). Having also in mind that,
\begin{equation}
\epsilon^{n}(\lambda) = {1 \over  \pi} {d \over d \lambda} \
p^{n}(\lambda)
\end{equation}
we end up with the following expression for the boundary
scattering amplitudes ($S_{\pm}^{n}=diag(\alpha_{\pm}^{n},
\beta_{\pm}^{n})$) \be \alpha_{+}^{n}\alpha_{-}^{n} = exp\Bigl \{
2piL\int_{0}^{\tilde \lambda_{1}}d\lambda \Bigl (\sigma(\lambda)
-\epsilon(\lambda)\Bigr ) \Bigr \} \ee and
 \be \alpha_{\pm}^{n}(\lambda, \xi^{\pm}) =  k_{0}^{n}(
\lambda)k_{1}^{n}(\lambda, \xi^{\pm}) \label{sol2} \ee where
\begin{equation}
k_{0}^{n}(\tilde \lambda_{1})= \exp \Bigl \{ \pi i
\int_{0}^{\tilde \lambda_{1}}\sum_{i=1}^{\nu}(K^{nn}(\lambda-
\tilde \lambda_{i})+K^{nn}(\lambda+ \tilde \lambda_{i}))
+Q^{n}(\lambda)
 d \lambda \Bigr \}
\end{equation} the $x$ dependent part is
\begin{equation}
k_{1}^{n}(\tilde \lambda_{1}, \xi^{\pm})= \exp \Bigl \{2 \pi i
\int_{0}^{\tilde \lambda_{1}} r^{n}(\lambda, \xi^{\pm})  d \lambda
\Bigr \}.
\end{equation}
We are interested in the limit that $q^{n} \rightarrow \infty$, in
this limit we can easily verify that \be \hat Q^{n}(\omega)
\rightarrow \hat K^{nn}(\omega)+{1 \over 2 \cosh({\omega \over
2})}, ~~\hat K^{nn}(\omega)  \rightarrow  {e^{-{\omega \over 2}}
\over 2 \cosh{\omega \over 2}} \ee
and
 \be
 \hat r^{1}(\omega, \xi^{\pm})
=-{e^{-(x^{\pm} -q^{1}){\omega \over 2}} \over 2 \cosh({\omega
\over 2})},~~\hat r^{2}(\omega,\xi^{\pm}) =0, \ee
 with $x^{\pm}-q^{1}$ to be a fixed number as $q^{n}
\rightarrow \infty$. 
Using the above Fourier transforms we end up with the following
expressions,   \be k_{0}^{n}(\lambda) &=& \exp \Bigl
\{2\int_{0}^{\infty} {d\omega \over \omega} \sinh (2i\omega
\lambda){\sinh({3\omega \over 2}) e^{-{\omega \over 2}} \over
\sinh(2\omega)} \Bigr \}, \non\\ k_{1}^{n}(\lambda, \tilde
\xi^{\pm}_{n}) &=& \exp \Bigl \{-\int_{0}^{\infty} {d\omega \over
\omega}\sinh(2i\omega \lambda){ e^{-2 \tilde \xi_{n}^{\pm} \omega}
\over \cosh(\omega)}\Bigr \}, \label{sol5b} \ee  where $2 \tilde
\xi_{1}^{\pm} =2\xi^{\pm} -2s^{1}+1$, $2 \tilde \xi_{2}^{\pm} \to
\infty $, are the renomarlized boundary parameters,
$k_{0}^{n}(\lambda)$ and $k_{1}^{n}(\lambda)$ are the $\xi$
independent and the $\xi$ dependent part, respectively of the
usual $XXX$ ($SU(2)$) reflection matrix (see e.g. \cite{gmn}).
 Exactly as in the case of the spin $s$ open chain, we consider the diagonal $K$ matrix, therefore we
 need to determine the other element of
 each boundary matrix. We exploit the ``duality'' symmetry
\cite{done} of the transfer matrix for $\xi^{\pm} \rightarrow
-\xi^{\pm}$, and we find that the other diagonal element is given
by \be \beta_{\pm}^{n}(\lambda, \tilde \xi_{n}^{\pm}) = e_{2
\tilde \xi_{n}^{\pm} -1}(\lambda) \alpha_{\pm}^{n}(\lambda, \tilde
\xi_{n}^{\pm}). \label{sol0b}\ee The ratio
${\beta_{\pm}^{n}(\lambda, \tilde \xi_{n}^{\pm}) \over
\alpha_{\pm}^{n}(\lambda, \tilde \xi_{n}^{\pm})}$ is the same as
in the $XXX$ model but with a renormalized boundary parameter. We
observe for the alternating chain as well as in the spin $s$ chain
that only one free boundary parameter $\tilde \xi^{\pm}_{1}$
($\tilde \xi^{\pm}_{2} \to \infty$) survives.

Two copies of the rational ($XXX$) reflection matrix were
computed, one for each excitation. Therefore, we conclude that the
boundary $S$ matrix should be of the structure, \be S(\lambda) =
S_{SU(2)}(\lambda, \tilde \xi_{1}^{\pm} ) \bigotimes
S_{SU(2)}(\lambda, \tilde \xi_{2}^{\pm}). \ee We assume that this
matrix should also coincide with the one of the $PCM_{1}$. A
calculation of the boundary $S$ matrix from the field theory point
of view would probably confirm our results. We have to mention
that there have been some studies for the $SU(2)$ $PCM$ with $WZW$
term with boundaries, \cite{feko} but mainly in the context of
quantum impurity (Kondo) problem. In particular, in \cite{feko}
the authors considered dynamical boundaries, i.e. they considered
quantum impurities at the boundaries, and they derived the
corresponding reflection matrices.

\section{Discussion}

The $XXX$ spin $s$ and the alternating $s^{1}$, $s^{2}$
($s^{1}-s^{2}={1\over 2}$) chains were explored. For both models
the Bethe ansatz equations were derived using fusion, and the
exact boundary $S$ matrices were computed. We were particularly
interested in the case that $s, s^{i}\rightarrow \infty $. More
specifically, for the spin $s \to \infty $ chain the boundary
scattering amplitudes were simply combined to give
 the boundary $S$ matrix of the form
  $f(\lambda)S_{SU(2)}(\lambda, \tilde \xi^{\pm}_{1} ) \bigotimes
 S_{SU(2)}(\lambda, \tilde \xi^{\pm}_{2})$. 

For the alternating spin chain two different types of excitations
exist: ${1\over 2}$ and 0 spin respectively. The boundary
scattering for each excitation was studied and the corresponding
reflection matrices were derived. Two copies of the $XXX$ boundary
$S$ matrix were computed ($s^{i}\rightarrow \infty $), and the
boundary $S$ matrix was given as a tensor product of two rational
boundary $S$ matrices. This matrix is also expected to coincide
with the one of the $SU(2)$ $PCM_{1}$. Note that we could end up
with to same result if we started from the anisotropic spin chains
and then take the isotropic and $s, s^{i} \to \infty$ limits.

It would be also interesting to consider dynamical $K$ matrices
\cite{feko} in order to construct the open spin chain and then to
study the reflection of the particle-like excitations with the
dynamical boundary. Another interesting aspect would be the study
of the thermodynamics of the alternating spin chain via the TBA.
The main purpose would be the derivation of the central charge
(see e.g. \cite{AM}, \cite{DMN1}), of the model for both bulk and
boundary cases. We hope to report on these issues in a future
publication.

\section{Acknowledgments}

I would like to thank  R.I. Nepomechie and E. Sklyanin for
valuable discussions and comments. Also, I am grateful
 to N. Reshetikhin for correspondence. This work is supported by
 the EPSRC.


\begin{thebibliography}{99}

\bibitem{bethe}
H. Bethe, Z. Phys. {\bf 71} (1931) 205.

\bibitem{FT2}
L.D. Faddeev and L.A. Takhtajan, Russ. Math. Surv. {\bf 34}, 11
(1979); L.D. Faddeev and L.A. Takhtajan, J. Sov. Math. {\bf 24}
(1984) 241.

\bibitem{FT1}
L.D. Faddeev and L.A. Takhtajan, Phys. Lett. {\bf 85A} (1981) 375.

\bibitem{R}
N.Yu. Reshetikhin, Nucl. Phys. {\bf B251} (1985) 565.

\bibitem{affl}
I. Affleck, {\it In fields strings and critical phenomena}, Les
Houches Lectures, (Elsevier 1990).

\bibitem{resh2}
N.Yu. Reshetikhin. J. Phys. {\bf A24} (1991) 3299.

\bibitem{cherednik}
I.V. Cherednik, Theor. Math. Phys. {\bf 61} (1984) 977.

\bibitem{sklyanin}
E.K. Sklyanin, J. Phys. {\bf A21} (1988) 2375; P.P. Kulish and
E.K. Sklyanin, J. Phys. {\bf A24} (1991) L435.

\bibitem{fk}
A. Fring and R. K$\ddot{o}$berle, Nucl. Phys. {\bf B421} (1994)
159; A. Fring and R. K$\ddot{o}$berle, Nucl. Phys. {\bf B419}
(1994) 647.

\bibitem{GZ}
S. Ghoshal and A. B. Zamolodchikov, Int. J. Mod. Phys. {\bf A9}
(1994) 3841; {\bf A9} (1994) 4353.

\bibitem{DVGR3}
H.J. de Vega and A. Gonz\'alez-Ruiz, Mod. Phys. Lett. {\bf A9}
(1994) 2207; H.J. de Vega and A. Gonz\'alez-Ruiz, J. Phys. {\bf
A26} (1993) L519.




\bibitem{baxter}
R.J. Baxter, Ann. Phys. {\bf 70} (1972) 193; J. Stat. Phys. {\it
8} (1973) 25; {\it Exactly Solved Models in Statistical Mechanics}
(Academic Press, 1982).

\bibitem{korepin}
V.E. Korepin, Theor. Math. Phys. {\bf 76} (1980) 165; V.E.
Korepin, G. Izergin and N.M. Bogoliubov, {\it Quantum Inverse
Scattering Method, Correlation Functions and Algebraic Bethe
Ansatz} (Cambridge University Press, 1993).


\bibitem{KR} A.Kirillov and
N.Reshetikhin, J. Sov. Math {\bf 35} (1986) 2621; A.Kirillov and
N.Reshetikhin, J. Phys. {\bf A20} (1987) 1565.

\bibitem{LT} L.A. Takhtajan, Phys. Lett. {\bf A87} (1982) 479.

\bibitem{PW} A.Polyakov and P.Wiegmann Phys. Lett., {\bf B131} (1983) 121.

\bibitem{W1} P.Wiegmann Phys. Lett. {\bf B142} (1984) 173.

\bibitem{fr1} L.D. Faddev and N.Yu. Reshetikhin, Ann. Phys. {\bf
167}(1986) 167.

\bibitem{B}
H.Babujian, Nucl. Phys. {\bf B215} (1983) 317.

\bibitem{DNW} G.Japaridze, A.Nersessian and P.Wiegmann,
Nucl. Phys. {\bf B230} (1984) 511.

\bibitem{BT} H.Babujian and A.Tsvelik, Nucl. Phys. {\bf
B265} (1986) 24.




\bibitem{ZAZ} A.Zamolodchikov and Al.Zamolodchikov, Nucl. Phys. {\bf
B379 (1992)} 602.

\bibitem{DB} D. Bernard, Phys.
Lett. {\bf B279} (1992) 78.

\bibitem{ZZ}
A.Zamolodchikov and Al.Zamolodchikov, Ann. Phys. {\bf 120} (1979)
253.

\bibitem{BR} V.V. Bazhanon and N.Yu.
Reshetikhin, Int. J. Mod. Phys. {\bf A4} (1989) 115.

\bibitem{abl}
C. Ahn, D. Bernard and A. Leclair, Nucl. Phys {\bf B346} (1990)
409; D. Bernard, A. Leclair,Phys. Lett. {\bf B247} (1990) 309.

\bibitem{VEWO} H.J. de Vega and F.
Woyanorovich, J. Phys. {\bf A25} (1992) 4499.

\bibitem{AM} S.R. Aladim and M.J.
Martins, J. Phys. {\bf A26} (1993) 7287.

\bibitem{DMN1} H.J. de Vega, L. Mezincescu and R.I. Nepomechie,
Phys. Rev. {\bf B49} (1994) 13223.

\bibitem{DMN} H.J de Vega, L.
Mezincescu and R.I. Nepomechie, Int. J. Mod. Phys. {\bf B8} (1994)
3473.

\bibitem{DM} B.D. Doerfel and S. Meisner, J. Phys. {\bf
A30} (1996) 6471.

\bibitem{bado}
A. Doikou and A. Babichenko, Phys. Lett {\bf B515} (2001) 220.

\bibitem{PW1}
A.Polyakov and P.Wiegmann, Phys. Lett. {\bf B141} (1984) 223.


\bibitem{mn/fusion}
L. Mezincescu and R.I. Nepomechie, J. Phys. {\bf A25} (1992) 2533.

\bibitem{nxxz}
R.I. Nepomechie, {\it hep-th/0110116}.

\bibitem{K} V.Korepin, Theor. Math. Phys {\bf 41} (1979)
53.

\bibitem{AD} N. Andrei and C. Destri, Nucl.
Phys. {\bf B131} (1984) 445.


















\bibitem{gmn}
M. Grisaru, L. Mezincescu and R.I. Nepomechie, J. Phys. {\bf A28}
(1995) 1027; A. Doikou, L. Mezincescu and R.I. Nepomechie, J.
Phys. {\bf A 30} (1997) L507.

\bibitem{done}
A. Doikou and R.I. Nepomechie, Nucl. Phys. {\bf B521} (1998) 547;
Nucl. Phys. {\bf B530} (1998) 641; Phys. Lett. {\bf B462} (1999)
321.

\bibitem{mash}
N.J. Makcay and B.J. Short, {\it hep-th/0104212}.

\bibitem{rsosb1}
C.Ahn and W.M. Koo, {\it hep-th/9708080}; J. Phys. {\bf A29}
(1996) 5845.

\bibitem{rsosb2}
R.E. Behrend, P.A. Pearce and D.L. O'Brien, J. Stat. Phys. {\bf
84} (1996) 1; R.E. Behrend and P.A. Pearce, J. Phys. {\bf A29}
(1996) 7827.

\bibitem{rsosb3}
M.T. Batchelor, V. Fridkin, A. Kuniba and Y.K. Zhou, Phys. Lett
{\bf B735} (1996) 266.



















\bibitem{feko}
P. Fendley, Phys. Rev Lett. {\bf 71} (1993) 2485; J.N. Prata,
Phys. Lett. {\bf B438} (1998) 115.

\end{thebibliography}
\end{document}